\documentstyle[aps,prd,twocolumn,epsfig]{revtex}
\newcommand{\be}{\begin{equation}}
\newcommand{\ee}{\end{equation}}
\newcommand{\beq}{\begin{eqnarray}}
\newcommand{\eeq}{\end{eqnarray}}

\begin{document}
\twocolumn[
\begin{center}
{\large\bf\ignorespaces
Probing hadron wave functions in Lattice QCD}\\
\bigskip
C.~Alexandrou \\
   {\small\it Department of Physics, University of Cyprus, P.O. Box 20537,
CY-1678 Nicosia, Cyprus} \\
\bigskip
Ph.~de~Forcrand \\
{\small\it Institute f\"ur Theoretische Physik, ETH H\"onggerberg,
 CH-8093 Z\"urich, Switzerland\\ and
CERN, Theory Division, CH-1211 Geneva 23, Switzerland}\\
\bigskip
A.~Tsapalis \\
 {\small\it Department of Physics, University of Wuppertal,
Wuppertal, Germany} \\
\medskip{\small\rm (June 20, 2002)} \\
\bigskip
\begin{minipage}{5.5 true in} \small\quad

Gauge-invariant equal-time correlation functions  are calculated
in lattice QCD within the quenched approximation and with 
two dynamical quark species. These correlators
provide information on the shape and multipole moments of the pion, the rho,
the nucleon and the $\Delta$.

\medskip
 
\end{minipage}
\end{center}
\vspace{6mm}
]

\section{Introduction}
State of the art lattice calculations of hadronic matrix elements 
have produced very accurate spectroscopic information.
Examples of the accuracy reached in quenched lattice QCD
are the calculation of the masses of low lying hadrons~\cite{CP-PACS}
 and glueballs~\cite{Morningstar}.
However  progress in determining hadron wave functions and quark distributions
has not been so rapid. The initial calculations of Bethe-Salpeter
amplitudes were carried out on a rather small lattice
in the mid 80's~\cite{VW} for the pion
and the rho. Further progress came in the early 90's when
the Bethe-Salpeter amplitudes were calculated on a larger lattice
for the pion and the rho~\cite{Hecht,Gupta0} as well as for the nucleon and the 
$\Delta$~\cite{Hecht}. However the results were of limited
interest because of their manifest dependence on the gauge chosen.
A different approach  to explore hadronic structure 
was pursued by the authors of refs.~\cite{Negele};
instead of fixing a gauge or a path for the gluons,
they considered correlation functions of quark densities
 which, being expectation values of local operators, are
gauge invariant.
This is the approach we have adopted in this  work.

 Our main motivation for studying density correlators is that 
 they reduce, in the non-relativistic limit,
 to the wave function squared and thus
they provide detailed, gauge-invariant information on hadron structure. The shape of hadrons
is one such important quantity that can be directly studied. 
The issue whether 
the nucleon is deformed from a spherical shape was raised twenty years ago~\cite{Isgur} and
is still unsettled.
Because the spectroscopic
quadrupole moment of a spin one-half particle  vanishes, 
in experimental studies one searches for quadrupole strength
in the  $\gamma^*N \> \rightarrow \> \Delta$ transition.
Spin-parity selection rules allow a magnetic dipole, M1, an electric
quadrupole, E2, or a Coulomb quadrupole, C2, amplitude. 
If both the nucleon and the $\Delta$
are spherical then the electric and Coulomb quadrupole amplitudes are
expected to be zero. Although M1 is indeed the dominant amplitude
there is mounting experimental evidence that E2 and C2 are non-zero.
The physical origin of a non-zero  E2 and C2 amplitude is
attributed to different mechanisms in the various models.
In quark models the deformation is due to the colour-magnetic tensor 
force\cite{Isgur}.
In ``cloudy'' baryon models it is due to meson exchange 
currents~\cite{Buchmann}.

A recent experimental search at $q^2=0.126$~GeV$^2$
has yielded
an electric quadrupole to magnetic dipole amplitude ratio~\cite{Mertz}:
\be
R_{EM}={\cal G}_{E2}/{\cal G}_{M1} = (-2.1\pm 0.2 \pm 2.0)\% \quad.
\ee
The larger error, an order of magnitude larger than the statistical error,
 is due to the model dependence in the extraction of
this ratio from the experimental data.
The experimental determination of $R_{EM}$ is complicated
 by the presence of nonresonant processes coherent with the resonant
excitation of $\Delta(1232)$.
 
The  ratio $R_{EM}$
 can be evaluated within lattice QCD without any model assumptions by
computing the transition matrix element $N$ to $\Delta$.
An early lattice calculation of this transition matrix element
provided  an estimate of $R_{EM}=(-3 \pm 8)\%$
~\cite{Leinweber}.

Here we consider an alternative route
to understanding the issue of deformation, via
the direct study of hadron wave functions.
We compute density-density correlators for mesons and baryons
and three-density correlators for baryons and look for
asymmetries when these are projected along the spin axis
or perpendicular to it.
 The observables that we use
are described in section II. 
In section III we give the relations for the deformation among the
states of different spin projections.
Our quenched lattice results are presented in section IV.
They clearly give support to a deformed rho. The nucleon
deformation averages to zero in agreement with the fact that its
spectroscopic quadrupole moment is zero. An analysis of the intrinsic 
nucleon deformation requires determining the body-fixed 
coordinates by diagonalization of the moment of inertia tensor, which
must be done 
configuration by configuration. This  is too noisy to yield a statistically
significant result. On the other hand the $\Delta$ has a non zero spectroscopic
moment and any deformation should be detected via projection with respect
to its spin axis. In  quenched QCD we detect no significant deformation
for the $\Delta$ within our present statistics.

In addition to providing information about deformation,
baryon wave functions, calculated for the first time in a gauge invariant
way for all values of the relative coordinates, 
can be used to study indirectly the potential among the three quarks. By 
performing fits to the
three-density correlators, we find that a reasonable ansatz for the baryonic potential
is provided by the sum of two-body potentials, called the $\Delta$-Ansatz~\cite{Cornwall}.

Comparison between quenched and full QCD results sheds light
on the role of the pion cloud.
It is known that quenching  eliminates
all or part of the pion cloud depending on the hadronic state.
For the rho channel no intermediate backgoing
quarks with the pion quantum numbers are present~\cite{pion}, so
the deformation that we observe is not due to the pion cloud. We use the SESAM
configurations~\cite{SESAM} to investigate unquenching effects. We find
that the deformation in the rho increases. We detect 
a deformation in the $\Delta$ which
remains small for the values of the dynamical quark mass considered. The unquenched
lattice results are presented in section V.
Section VI contains our conclusions.

\section{Correlation functions}

The wave function of a meson is usually defined in a given gauge $g$ 
as the equal time Bethe-Salpeter amplitude
\be
\Phi^{\rm BS}_g({\bf r})=\int d^3r' 
     <0|\bar{q}^{ f_1}({\bf r}')\Gamma q^{f_2}({\bf r}'+{\bf r})|M>
\ee
where $\Gamma$ is a Dirac matrix with the quantum numbers of the
meson $M$ of flavor $f_1, f_2$.
$\Phi^{\rm BS}_g({\bf r})$ is 
the minimal Fock space state wave function, which
is an approximation to the full wave function since 
other multiquark components are excluded.
For the Bethe-Salpeter amplitude  one must either fix the gauge
or connect the quarks with gluons to form gauge-invariant 
(but path-dependent) quantities.
Wave functions for baryons are defined in an analogous 
way to the  meson wave functions but
they involve two relative distances.
E.g. the Bethe-Salpeter amplitude, $\Phi^{\rm BS}_g({\bf r}_1,{\bf r}_2)$, 
for the proton is given by
\be
\sum_{\bf r'} <0| \epsilon^{abc}
  u^a_{\delta}({\bf r}',t)
\biggl(u^{b\>T}({\bf r}'+{\bf r}_1,t) \>C\gamma_5 \>d^c({\bf r}'+{\bf r}_2,t)\biggr)
 |B> \quad . 
\ee
Summation over ${\bf r}'$ projects onto zero momentum.
For the $\Delta^+$, the interpolating field that we take is

\beq
J_\mu({\bf x})&=&\frac{\epsilon^{abc}}{\sqrt{3}}
\Biggl[u^a({\bf x})\biggl(2u^{b\>T}({\bf x})\>C\gamma_\mu d^c({\bf x})\biggr)\nonumber \\
&+&d^a({\bf x})\biggl(u^{b\>T}({\bf x})\>C\gamma_\mu u^c({\bf x}) \biggr)\Biggl] \quad ,
\label{delta}
\eeq
where $C=\gamma_0\gamma_2$ is the charge conjugation operator.
We use  $\Gamma_{\pm}=(\gamma_1\mp i\gamma_2)/2$ to 
create a hadron of  a definite spin component $J_z$.
Explicitly for the first term in Eq.~\ref{delta} we take 
\beq
J_{3/2} &:& \frac{\epsilon^{abc}}{\sqrt{3}}
\Biggl[u^a_1({\bf x})\biggl(2u^{b\>T}({\bf x})\>C\Gamma_{+} d^c({\bf x})\biggr)\Biggr]\nonumber \\ 
J_{1/2} &:& \frac{\epsilon^{abc}}{\sqrt{3}}
\Biggl[u^a_1({\bf x})\biggl(2u^{b\>T}({\bf x})\>C\gamma_3 d^c({\bf x})\biggr)\nonumber \\ 
& & - u^a_2({\bf x})\biggl(2u^{b\>T}({\bf x})\>C\Gamma_{+} d^c({\bf x})\biggr)\Biggr]\nonumber \\ 
J_{-1/2} &:& \frac{\epsilon^{abc}}{\sqrt{3}}
\Biggl[u^a_2({\bf x})\biggl(2u^{b\>T}({\bf x})\>C\gamma_3 d^c({\bf x})\biggr)\nonumber \\ 
& & + u^a_1({\bf x})\biggl(2u^{b\>T}({\bf x})\>C\Gamma_{-} d^c({\bf x})\biggr)\Biggr]\nonumber \\
J_{-3/2} &:& \frac{\epsilon^{abc}}{\sqrt{3}}
\Biggl[u^a_2({\bf x})\biggl(2u^{b\>T}({\bf x})\>C\Gamma_{-} d^c({\bf x})\biggr)\Biggr]
\quad .
\label{spin projections}
\eeq
where $q_i$ is the $i$th component of the spinor. 
The same construction is done for the second term of Eq.~\ref{delta}.
With these combinations we recover, in the non-relativistic limit, the 
quark model wave functions of these states.

The Bethe-Salpeter amplitudes in the Coulomb and in the Landau gauge
were investigated in ref.~\cite{Hecht}.
Instead of fixing a gauge,
a gauge-invariant wave function, which corresponds to
the Bethe-Salpeter amplitude in the axial gauge, can be constructed  
by joining the quark 
and the antiquark with a thin string of glue~\cite{Negele3}:
\begin{equation}
 \psi_s(y) = \bigl\langle0\bigl| \bar{q}(0)\Gamma e^{i\int^{y}_{0} dz A_y(z)}
 q(y) \bigr|M\bigr\rangle \quad .
\label{eq:thin path}
\end{equation}
Other variations to the thin string
are to smear the gluons~\cite{Gupta0} or to evolve  
them so that they reach their ground state distribution. 
It was shown in ref.~\cite{Negele3} that  there is a much larger probability
to find a quark-antiquark pair separated by, say, 1 fm when connected by a
physical adiabatic flux tube than when the quarks are surrounded
by gluons fixed to the Coulomb gauge or when they are connected by a
thin string of gluons. Although in ref.~\cite{Negele3} 
only mesonic states were considered, generalization to baryons is 
straightforward, with the gluonic strings attached to each of the three quarks 
contracted at a point.

In this work we opt for the calculation
of density-density correlators~\cite{Negele2,Wilcox},

\be
 C({\bf r},t_1,t_2) = \int\> d^3r'\>
\langle h|\rho^{u}({\bf r}'+{\bf r},t_2)\rho^{d}({\bf r}',t_1)|h\rangle 
\label{2density}
\ee
where the density operator is given by the normal order product
\be
\rho^u({\bf r},t) = :\bar{u}({\bf r},t)\gamma_0 u({\bf r},t): 
\ee
so that disconnected graphs are excluded.

\begin{figure}[h]
\begin{center}
\epsfxsize=8.0truecm
\epsfysize=5.truecm
\mbox{\epsfbox{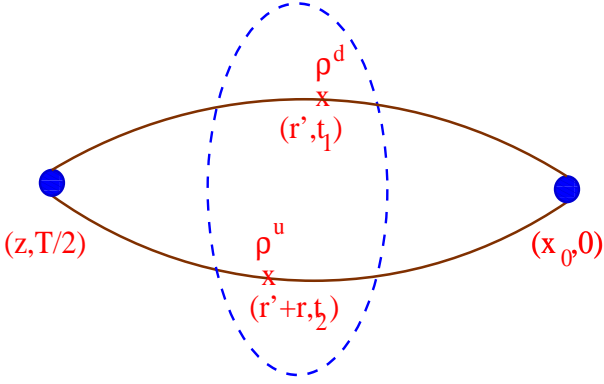}}
\vspace*{0.5cm}
\caption{Density-density correlator for a meson. $t_1$, $t_2$,
 $T/2-t_1$ and $T/2-t_2$
are taken large enough to isolate the mesonic groundstate.} 
\end{center}
\label{fig:2density}
\end{figure}

For mesons the density-density correlator is shown schematically in 
 Fig.~\ref{fig:2density}. 
Inserting a complete set of hadronic states we find
\beq
 C({\bf r},t_1,t_2)& =& \sum_{n_i,n_f,n}\sum_{\bf{p},\bf{q}}
\langle h|n_f,{\bf{p}}\rangle 
\frac{e^{-E_{n_f}({\bf p})(T/2-t_2)}}{E_{n_f}({\bf p})} \nonumber \\
&\>&\langle n_f,{\bf p}|\rho^u|n,{\bf p}+{\bf q}\rangle 
e^{-i{\bf q.r}}\frac{e^{-E_n({\bf p}+{\bf q})(t_2-t_1)}}{E_n({\bf p}+{\bf q})}
\nonumber \\
&\>& \langle n,{{\bf p}+{\bf q}}|\rho^d|n_i,{{\bf p}}\rangle 
\frac{e^{-E_{n_i}({\bf p})t_1}}{E_{n_i}({\bf p})} \langle n_i,{\bf p}|h\rangle
 \quad.
\eeq
The sum over all excitations of the initial, $n_i$, and 
final, $n_f$, states  yields the ground state hadron in the
limit $(T/2-t_2)\rightarrow \infty$ and $t_1 \rightarrow \infty$,
where $T$ is the lattice extent in Euclidean time. 
Since we take periodic boundary conditions
the maximum time separation is $T/2$. 
Enforcing
zero momentum would require a summation over the spatial volume
on the source or sink site. This is technically not feasible since
it involves quark propagators from all to all spatial lattice sites.
Instead, the suppression of the non-zero momenta and other higher excitations 
is obtained by choosing the
largest possible time separation from the source and the sink.
We  take  the  same spatial coordinate for the source and the sink, 
namely ${\bf x}_0 = {\bf z}$.
As a starting point we take, in this work,   density
insertions to be always at equal times.         
A disadvantage
of the density-density correlators  is that
they are subject to more severe finite size effects, having
typically larger spatial extent ($\sim$ twice)
 than Bethe-Salpeter amplitudes~\cite{Negele}.

In the case of baryons 
three density insertions are needed and the correlator is given by 
\be
 C({\bf r}_1,{\bf r}_2,t) = \int d^3r'\>
\langle h|\rho^{d}({\bf r'},t)\rho^{u}({\bf r}'+{\bf r}_1,t)\rho^{u}({\bf r}'+{\bf r}_2,t)|h\rangle
\label{3density}
\ee
where we have taken all the density insertions to be at equal times. 
This involves two relative distances.
Like (\ref{2density}), it can be computed efficiently by Fast Fourier Transform (FFT).

\begin{figure}[h]
\epsfxsize=8.0truecm
\epsfysize=8.truecm
\mbox{\epsfbox{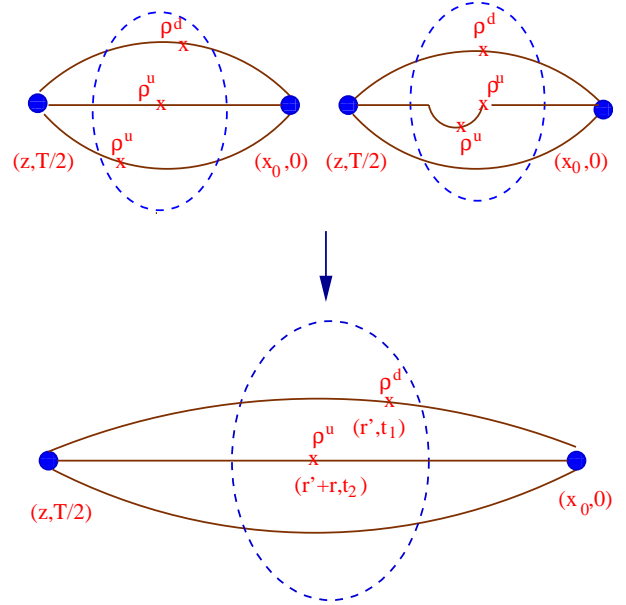}}
\vspace*{0.5cm}
\caption{The upper two diagrams show the three-density correlator 
for a baryon as defined in Eq.~\ref{3density}. As in Fig.~1 the density
operators are inserted far enough from the source and the sink so that
the baryonic groundstate of interest is isolated. The lower diagram shows
the equivalent two-density correlator after integration of one of the relative
distances.}
\label{fig:3density}
\end{figure}

The 
relevant diagrams are shown in the upper part of
Fig.~\ref{fig:3density} for the nucleon or the $\Delta^+$,
for the general case of density insertions at three unequal 
times.
In addition to one density insertion for each $u$-quark line, one 
can have two density insertions on the same $u$-quark line. Evaluation of
this second diagram requires the quark propagator $G({\bf r}_2,{\bf r}_1)$ for
all partial distances of the two arguments. This is beyond 
 our present resources, and this diagram is not included.
To check that the first diagram that we calculate provides by itself a reasonable
description of the baryon wave function, we also compute the baryon
wave function with two density insertions for the $u$ and $d$ quarks as
shown in the lower part of Fig.~\ref{fig:3density}. This may be viewed
as the square of the one-particle wave function obtained from the
full wavefunction by integrating over one relative coordinate.
If the contribution of the second diagram is small then we expect that 
\be
\int d^3 r_2 C({\bf r}_1,{\bf r}_2,t) \sim
\int\> d^3r'\>\langle h|\rho^{d}({\bf r}',t)\rho^{u}({\bf r}'+{\bf r}_1,t)|h\rangle
\label{oned}
\ee
will be satisfied, even when
the l.h.s of the equation is calculated using only the diagram
with one density insertion on each quark line.
This comparison will be performed in Sec.IV, B.

\section{Relations among the deformations in different channels}
The interpolating field for a rho meson is 
taken to be $J_\mu(x)=\bar{d}(x)\gamma_\mu u(x)$. 
The physical states of spin projections
$0$ and $\pm 1$ for the rho are obtained using interpolating fields
$J_0(x) = \bar{d}(x)\gamma_3 u(x)$ and 
$J_{\pm}= \bar{d}(x)\left[(\gamma_1\mp i \gamma_2)/2\right] u(x)$ 
respectively.
Due to rotational invariance   the correlator  
\be
C_{ss}({\bf r}) = \sum_{{\bf x}} \langle J_s({\bf z})\rho^d({\bf x})
\>\rho^u({\bf x+r})J_s^{\dagger}({\bf 0}) \rangle
\ee
satisfies the relation
$C_{11}({\bf r.\hat{e}_1})=C_{22}({\bf r.\hat{e}_2})=C_{33}({\bf r.\hat{e}_3})$
where $\hat{\bf e}_j$ is a unit vector along the $j$-axis. 
This means that for the $\pm 1$ channels  we have
$(C_{--}+C_{++})/2 =  C_{11}+C_{22}$. 
Due to parity symmetry 
$C_{--}= C_{++}$ which is satisfied after ensemble averaging and leads to
the relations
\be
C_{++}= C_{--}= C_{11}+C_{22} \quad .
\ee

\begin{figure}[h]
\begin{center}
\vspace*{-2.5cm}
\begin{minipage}{8cm}
\epsfxsize=7.truecm
\epsfysize=7.truecm
\mbox{\epsfbox{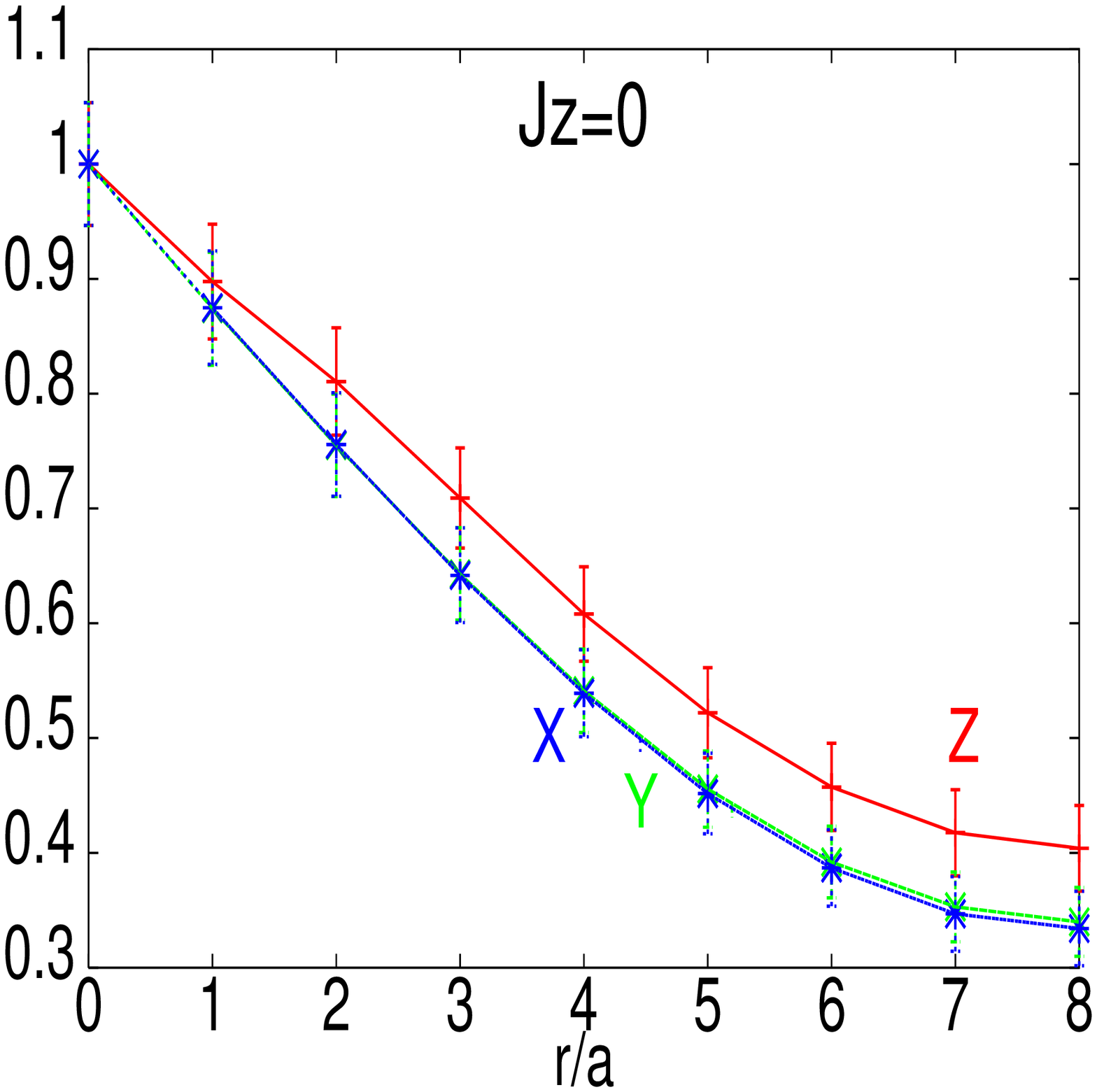}}
\end{minipage}
\begin{minipage}{8cm}
\vspace*{-2cm}
\epsfxsize=7.truecm
\epsfysize=7.truecm
\mbox{\epsfbox{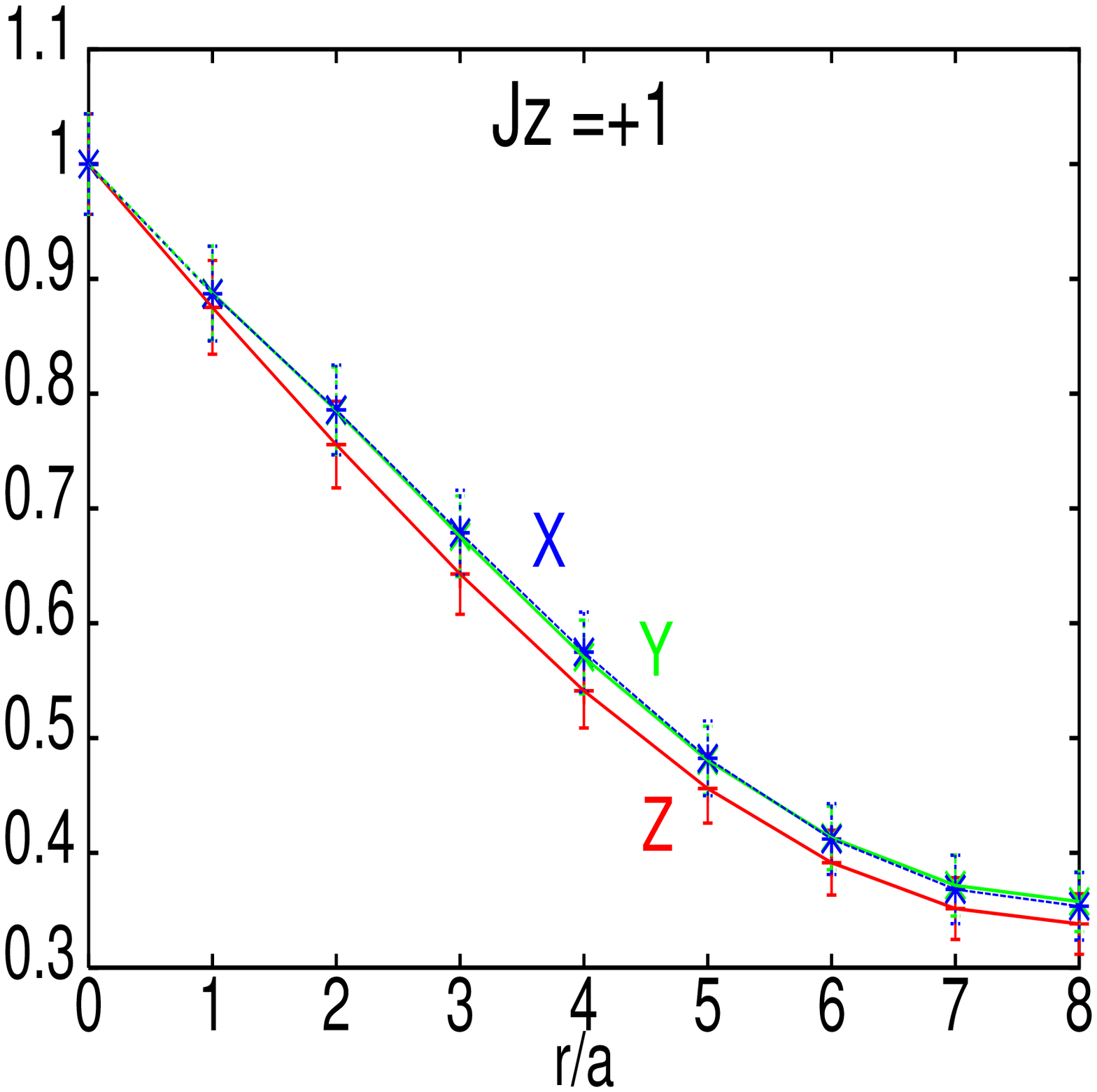}}
\end{minipage}
\vspace*{0.5cm}
\caption{
Density-density correlator for the rho in a definite spin state  
$J_z = 0$ (top) and $J_z = +1$ (bottom), for quark 
separations transverse (line with Z label)
and longitudinal (line with X and  Y labels) with respect to the spin axis. The hopping parameter is $\kappa=0.154$.}
\label{fig:rho-spin}
\end{center}
\end{figure}

If we denote by $T(r)$ the tranverse 
and by  $L(r)$ the longitudinal projection of $C_{33}(r)$ 
with respect to  the spin axis, 
the deformation $\alpha(r)$
is defined by
$L(r)/T(r)=1+\alpha(r)$. The deformation in the $\pm 1$ channels is
then $2T(r)/(T(r)+L(r)) \sim 1-\alpha(r)/2$ for small $\alpha$ i.e 
if the spin-0 state of the rho
 is elongated (prolate) along the $z$-axis then the spin $\pm 1$ states  will be 
``flat'' (oblate) by approximately half this amount. This observation is consistent
with the data shown in Fig.~\ref{fig:rho-spin}.

Non-relativistically one can use the Wigner-Eckart theorem to obtain
a similar result. 
The deformation is obtained by measuring
the quadrupole moment defined by
\beq
Q &=& \langle JM| 2(z^2-\frac{x^2+y^2}{2}) |JM\rangle \nonumber \\
&=&\sqrt{\frac{4\pi}{5}}\langle JM| 2 r^2 Y_{20}(\theta,\phi) |JM\rangle \nonumber \\
&=& (-)^{J-M}
  {J \>\> 2 \> J \choose -M \> 0 \> M} 
\langle J|2 r^2 P_{2}(cos(\theta)) |J\rangle 
\label{Wigner-Eckart}
\eeq
for a state $|JM\rangle$ where J is the spin of the state, M the
spin projection along the $z$-axis 
and $ {J \> \> 2 \> \>J \choose -M \> 0 \> M} $ 
 is the 3-j symbol. Therefore if we know the deformation for
one spin projection we can relate it to the rest. Evaluating the 3-j
symbol for the quantum numbers of the rho we find that
the deformation for $M=0$ is twice and opposite in sign to that
for $M=\pm 1$. We will thus only show results for the spin-0 state, which
exhibits the largest deformation. 

The interpolating fields for projecting
 to the physical $\Delta^+$ spin states  were given in 
Eq.~\ref{delta}.
Just like for the deformation of the rho in different spin projections,
similar relations can be obtained for the physical components of the 
$\Delta^+$.
The interpolating fields for positive spin projections $3/2$ and
$1/2$ are no longer related to the negative ones since they involve 
different spinor components. 
However the cross terms 
of the type $\gamma_1\gamma_2$
contribute an order
of magnitude less to the correlator and if we neglect these terms
we 
find for the density-density correlator of the $\pm 3/2$  state that
$C_{3/2} \sim C_{-3/2} \sim 1-\alpha(r)/2$ \footnote{The small
difference between $C_{3/2}$ and  $C_{-3/2}$ 
is due to our limited statistics.} where
$\alpha(r)$ is the deformation if we take $\gamma_3$ in Eq.~\ref{delta}.

A relation among all the spin projections  of the $\Delta^+$
 can be obtained in the non-relativistic
limit by
applying the Wigner-Eckart theorem Eq.~\ref{Wigner-Eckart}.
We find that the deformation  for the  $M=\pm 3/2$ states is equal and
opposite to that for the $M=\pm 1/2$. Among the physical states
we will therefore show results
 only for the $+3/2$.
Since we expect the deformation to be maximal
when a $\gamma_3$ is used in  Eq.~\ref{delta} for
the interpolating field, we will also look for deformation
in this channel in addition to the $+3/2$ physical state.
In the unquenched data where we observe a small deformation these
relations between the various amplitudes, as far as the relative signs
are concerned, are indeed satisfied.

\section{Quenched Lattice results}
\subsection{Density-density correlation functions}
We have analysed 220  quenched configurations at $\beta=6.0$
for   a lattice of size $16^3 \times 32$ obtained 
from the NERSC archive~\cite{NERSC}, using the Wilson Dirac operator with
hopping parameter $\kappa=0.15, 0.153, 0.154$ and $ 0.155$. 
The ratio of the pion mass to the rho mass at these values of $\kappa$  is
$0.88,\>0.84,\>0.78$ and 0.70 respectively. Using the relation
$2am_q=1/\kappa-1/\kappa_c$, with the critical value $\kappa_c=0.1571$,
we obtain for the naive quark mass $m_q$  values of about 
300, 170, 130 and 90~MeV respectively, where we used $a^{-1}=1.94$~GeV 
($a=0.103$ fm) from the string tension~\cite{Bali} 
 to set the scale.
Alternatively, the scale could be set from
the rho mass in the chiral limit. This approach yields $a^{-1}$=2.3~GeV 
($a=0.087$ fm)~\cite{QCDPAX,Gupta}, with a systematic error of about 10\%
coming from the choice of fitting range and chiral extrapolation ansatz,
which is about twice as large as the statistical one~\cite{QCDPAX}.
We note that although that these simulations used the lattices of
increasing bigger volumes the scale was not affected.
In our discussion of 
quenched data we will use the value of $a$ determined from the string tension. 
However, to compare the quenched  with the unquenched results we will
use the value extracted from the rho mass in the chiral limit since this
determination is applicable both in the quenched and in the unquenched theory.

We  fix the source and the sink for maximum separation
at $t_i=a$ and $t_f=17a$,
where $a$ is the lattice spacing. The
density insertions are taken  in the middle of the
time interval i.e. at $t=9a$. 
To check that the time interval $|t - t_i| = |t - t_f| = 8a$ is sufficient,  
we have performed an analysis on 27 configurations at $\kappa=0.153$,
varying the time $t$ where the density operators are inserted.

\begin{figure}[h]
\epsfxsize=8.0truecm
\epsfysize=9.5truecm
\mbox{\epsfbox{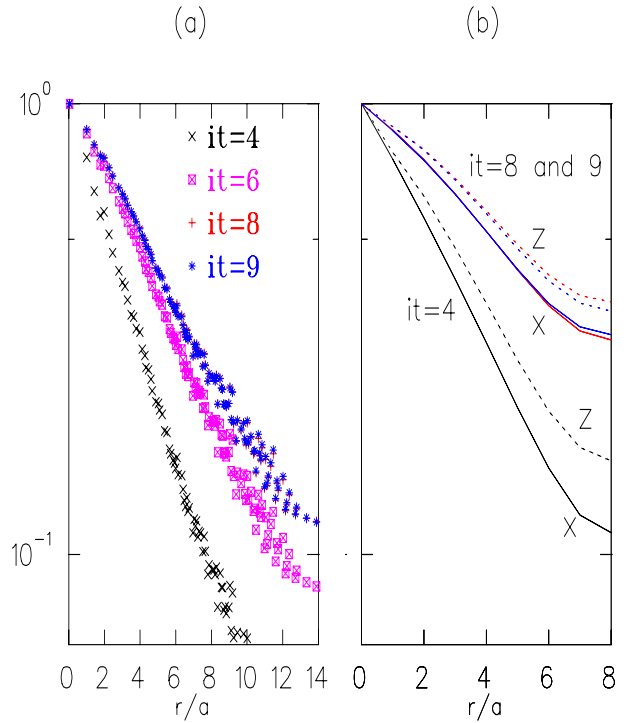}}
\vspace*{0.5cm}
\caption{ (a) The density-density correlator for the rho, measured from
density insertions at time $9a, 8a, 6a $ and $4a$. 
To avoid cluttering, the data as a function of $r$ are averaged over bins of size $\Delta r = 0.07a$. 
(b) The density-density correlator for the rho, measured from 
density insertions at time $9a, 8a $ and $4a$, for quark separations along the $z$- and
$x$-axes. In both cases, 27 configurations are used, and statistical error bars
are omitted for clarity.}
\label{fig:rho_it}
\end{figure}

 The results for the rho correlator
 are shown in 
Fig.~\ref{fig:rho_it}(a) for four
different insertion times at $\kappa=0.153$. 
As can be seen, density insertions at $t=8a$ and $9a$ give the same correlator,
reassuring us that the time separation is large enough to isolate the 
groundstate of the rho. In the other channels the results are similar,
with larger statistical noise for the baryons.
In  
Fig.~\ref{fig:rho_it}(b) we show in addition the 
correlator for the spin-0 state of the rho, for quark separations
 along the spin axis ($z$)  and perpendicular to it ($x$).
As expected, the deformation is the same when measured from density insertions
at $t=8a$ and $t=9a$, since the groundstate is isolated by then.
More surprisingly, this $z-x$ asymmetry is almost unchanged when measured 
at $t=4a$, even though the $z$- and $x$-profiles change appreciably.
These findings indicate that the deformation which we observe in more detail
below is a robust, 
physical property of the rho meson in its groundstate as well as its
low-lying excited states.

\begin{figure}[h]
\epsfxsize=8.0truecm
\epsfysize=17.5truecm
\mbox{\epsfbox{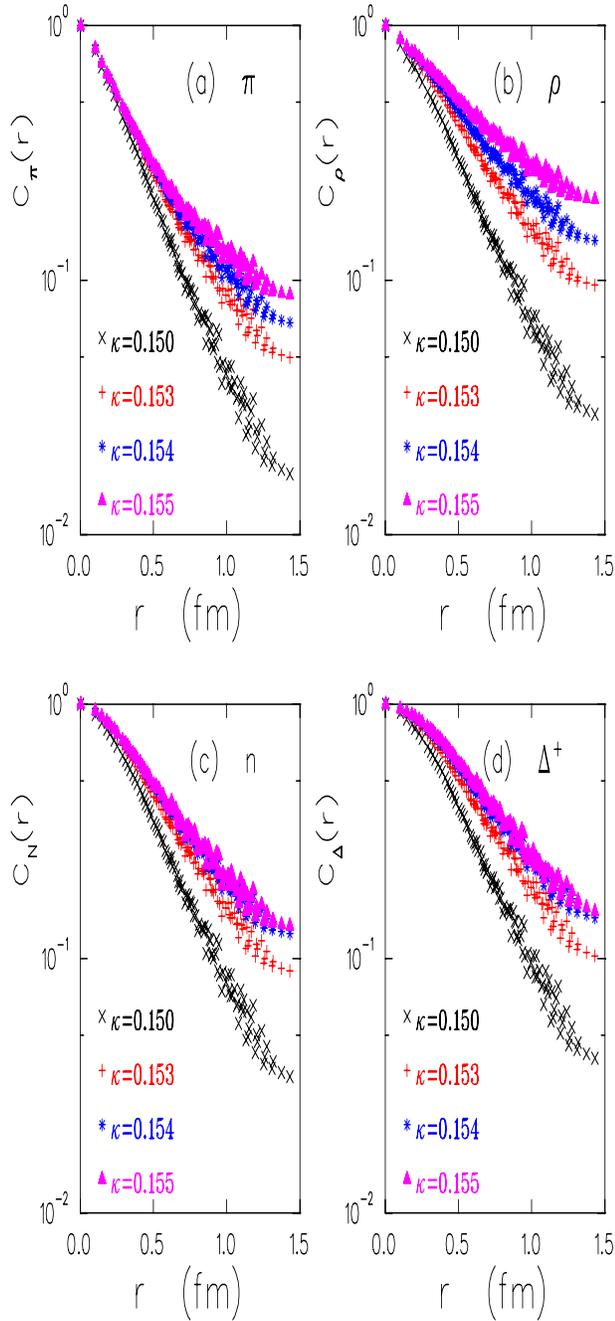}}
\vspace*{0.5cm}
\caption{Density-density correlators, $ C({\bf r})$, for
(a) the pion, (b) the rho, (c) the nucleon and
(d) the $\Delta^+$  versus $|{\bf r}|$ at $\kappa=0.15, 0.153, 0.154$ and $0.155$. 
Errors bars are omitted for clarity.}
\label{fig:all wfs}
\end{figure}

In Fig.~\ref{fig:all wfs} we collect the correlators for the pion, the rho,
the nucleon and the $\Delta^+$ for the  four
different quark masses ($\kappa$ values) considered.
We confirm an observation made in earlier studies~\cite{VW,Hecht} that 
the wave functions 
are not very sensitive to the bare quark mass. This behaviour
is as expected in the bag model,
 but is inconsistent with nonrelativistic quark models where a much stronger
mass dependence is predicted. Insensitivity on the bare mass
 can be understood from the consideration that
the quarks are dressed and their effective mass is 
therefore not very much affected
by changing the bare mass.
 From Fig.~\ref{fig:all wfs} we also 
see that the rho wavefunction depends more on the quark mass than
the other three. In particular the nucleon and $\Delta^+$ wave 
functions
hardly change as we go from $\kappa=0.154$ to
 $\kappa=0.155$ which corresponds to reducing the  naive
quark mass from $m_q\sim 130$~MeV to $m_q\sim 90$~MeV.

\begin{figure}[h]
\epsfxsize=8.0truecm
\epsfysize=8.truecm
\mbox{\epsfbox{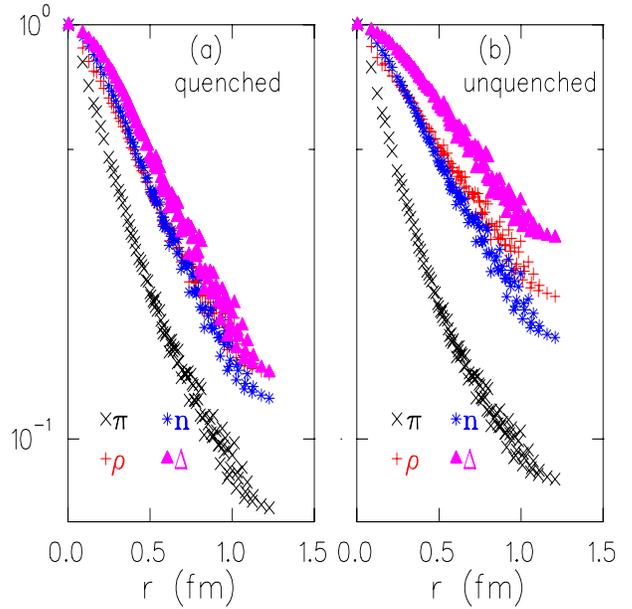}}
\vspace*{0.5cm}
\caption{ (a) Density-density correlators, $ C({\bf r})$,
for the pion, the rho, the nucleon and
the $\Delta^+$ at $\kappa=0.154$ vs $|{\bf r}|$. 
(b) Same as (a) but with two dynamical quarks at $\kappa=0.157$.
The dynamical results will be discussed in section V.
We used the rho mass to set the scale in both the quenched and the unquenched
theory.
Errors bars are omitted for clarity.}
\label{fig:wfs}
\end{figure}

A direct comparison of the sizes of
the four hadrons can be made 
in Fig.~\ref{fig:wfs}(a) where we plot the density-density correlators
for $\kappa=0.154$.
The pion has the smallest size, approximately half that of the $\Delta$,
whereas the rho has a size comparable to the nucleon and the $\Delta$.

\begin{figure}[h]
\vspace*{-0.5cm}
\epsfxsize=11.5truecm
\epsfysize=8truecm
\mbox{\epsfbox{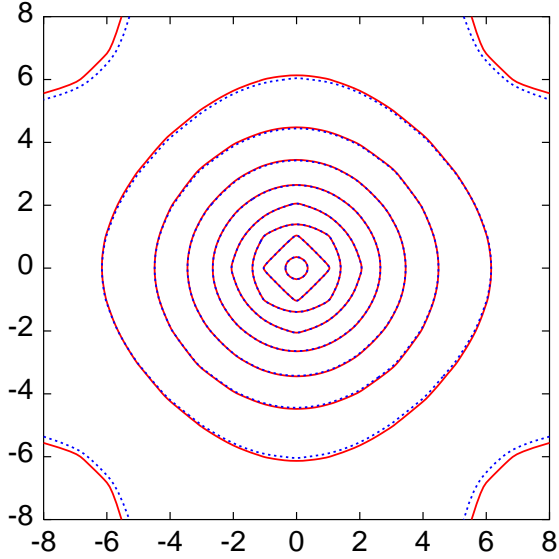}}
\vspace*{-0.5cm}
\caption{Contour plot of the pion correlator, $C({\bf r})$,
at $\kappa=0.153$, when $r$ lies in the $xz$-plane (solid lines)
or in the $xy$-plane (dashed lines) of size $16^2$.}
\label{fig:contour-pion}
\end{figure}

\begin{figure}[h]
\vspace*{-0.5cm}
\epsfxsize=11.5truecm
\epsfysize=8truecm
\mbox{\epsfbox{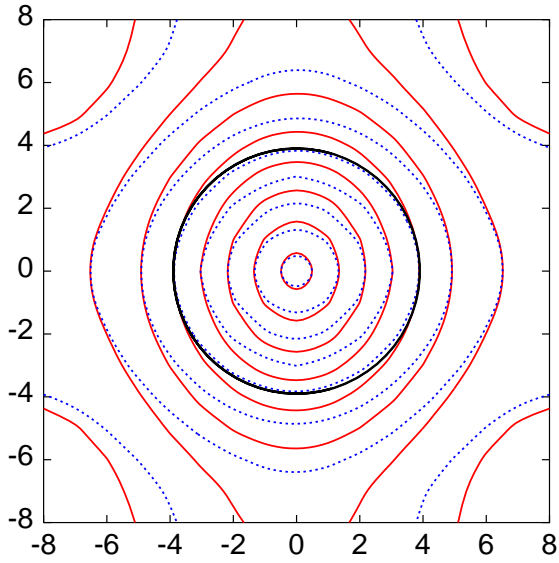}}
\caption{Same as Fig.~\ref{fig:contour-pion} but for the rho for $J_z=0$.
We have included a circle to guide the eye.}
\label{fig:contour-rho}
\end{figure}

\begin{figure}[h]
\epsfxsize=11.5truecm
\epsfysize=8truecm
\mbox{\epsfbox{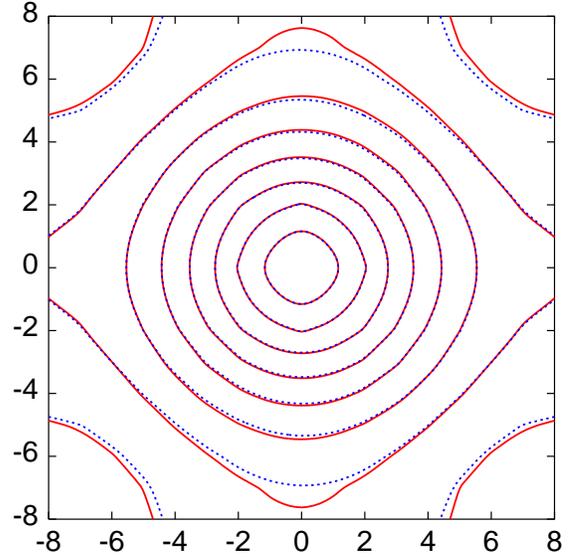}}
\caption{Same as Fig.~\ref{fig:contour-pion} but for the nucleon.}
\label{fig:contour-nucleon}
\end{figure}

\begin{figure}[h]
\vspace*{-0.5cm}
\epsfxsize=11.5truecm
\epsfysize=8truecm
\mbox{\epsfbox{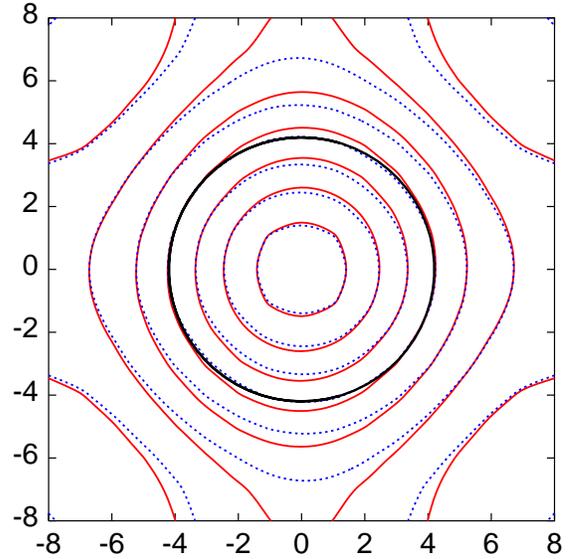}}
\caption{Same as Fig.~\ref{fig:contour-rho} but  for the $\Delta^+$.
The interpolating field is $J_3$.}
\label{fig:contour-delta}
\end{figure}

Any asymmetry with respect to the spin axis $z$ is best seen 
by comparing the correlator $C({\bf r})$ for ${\bf r}=(x,y,0)$ and $(x,0,z)$, 
i.e. in  a plane perpendicular
to the spin axis and a plane containing it.
The resulting contour plots are shown for $\kappa=0.154$
in
Figs.~\ref{fig:contour-pion} and~\ref{fig:contour-rho} 
for the pion and the rho
 and in Figs.~\ref{fig:contour-nucleon} and~\ref{fig:contour-delta} 
for the nucleon and the $\Delta^+$.
 The cigar shape
is clearly visible in the case of the rho, whereas the pion and the nucleon
 look spherical (up to small lattice distortions)
as expected since their spectroscopic quadrupole is zero.

A small asymmetry appears for the $\Delta^+$, but it is not statistically
significant. 
This is true for all the $\Delta^+$ spin projections. 
In Fig.~\ref{fig:contour-delta} 
we select the unphysical state obtained using interpolating field $J_3$
because it should show maximal deformation. In the case of the $\Delta$,
quenching removes only part of the pion
cloud, contrary to the case of the rho where it is removed completely.
Still, our negative finding does not rule out a deformation of the $\Delta$
induced by pions.  
It may be that
the asymmetry  only shows up when the quark mass is further decreased
so that the pion has a mass close to its physical value, although
we do not observe a statistically significant increase in the
deformation as we go from quenched quarks of naive quark mass 
$m_q\sim 300$~MeV to $m_q\sim 90$~MeV. 
It could also be that the asymmetry is enhanced and only becomes visible 
in full QCD, with a complete pion cloud made of light enough pions.
This issue is addressed in the next section.

\begin{figure}[h]
\epsfxsize=8.0truecm
\epsfysize=8.5truecm
\mbox{\epsfbox{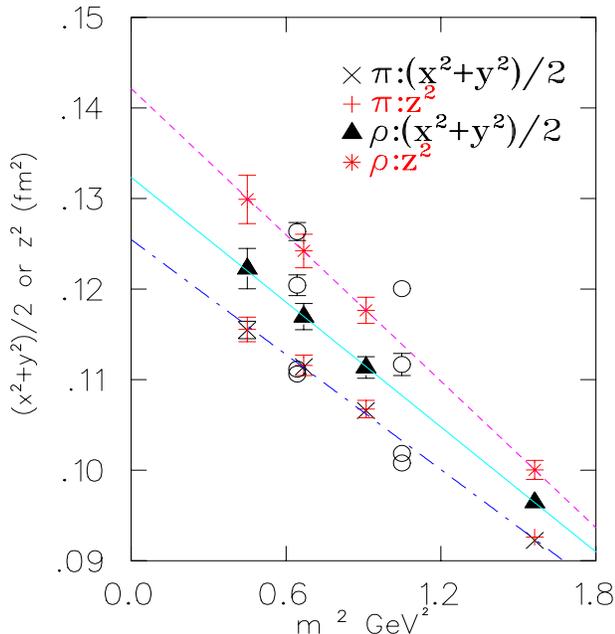}}
\caption{$\langle z^2 \rangle $ and $\langle (x^2+y^2)/2 \rangle $ versus
the pion mass squared in physical units. The lines are linear fits to the
quenched data.  
The pion (dash-dotted line) is spherical. It is  smaller than the rho,
whose transverse size with respect to the spin axis (solid line) is
smaller than its longitudinal size (dashed line). The rho
(in the  spin-0 projection) is shown to be cigar shaped, particularly in the chiral limit.
The circles show full QCD results, discussed in section V.}
\label{fig:meson-radii}
\end{figure}

A more quantitative determination of meson deformations can be obtained
 by computing the second
moments of the quark separation ${\bf r}$ along the three axes. 
Fig.~\ref{fig:meson-radii} shows the second  moments along the spin-axis, 
$\langle z^2 \rangle $, 
and the average of the moments along the two transverse axes, 
$\langle (x^2+y^2)/2 \rangle $, plotted versus
the pion mass squared. Here we use the rho mass to convert to physical 
units in order to be able to compare with the corresponding unequenched 
results.

The spin-0 state of the rho shows an elongation  along
the spin-axis, which increases as the quark mass is decreased. As already mentioned, since
this is a quenched calculation, this deformation is not due to the
pion cloud. Therefore the situation may change if
dynamical quarks are included. This is studied in Section V.

From the second moments we can obtain the charge root mean square (rms) radius
of the mesons defined in the quark model by 
\beq
<r^2_{\rm ch}> &=&  \sum_q e_q \langle({\bf r}_q- {\bf R})^2\rangle
 \nonumber \\
&=& \frac{\sum_q e_q\int d^3r\> ({\bf r}/2)^2\> C({\bf r})}{ \int d^3r\>C({\bf r})}
\label{rms radius}
\eeq
where ${\bf R}$ is the coordinate of the
 center of mass and $e_q$ is the electric charge of the quarks. In the chiral (quenched) limit
we estimate for the pion $\sqrt{r^2_\pi} \sim 0.35$~fm using the rho mass
to set the scale and 0.42~fm using the string tension~\cite{Bali}
 to be compared
with the experimental value of 0.53~fm~\cite{Amendolia}. 
For the rho, using the rho mass and the string tension  we find respectively
 $\sqrt{r^2_\rho} \sim 0.37$~fm  and 0.44~fm.
The ratio $\sqrt{r^2_\rho/r^2_\pi}\sim 1.06$ from the lattice data is  
somewhat small
 compared  to the value 1.15, obtained from the experimental value 0.53~fm for the 
pion~\cite{Amendolia} and 
 $0.61$~fm~ for the rho~\cite{Hawes} calculated
in a Dyson-Schwinger equation approach using $f_\pi=93$~MeV.
 This discrepancy
between the experimental and lattice results  may be due to two reasons:
1) we are using 
the quenched theory where the pion cloud is eliminated and 
2) we are far from the chiral limit and making a linear extrapolation
in $m_\pi^2$ may be problematic, 
especially since chiral loops give a logarithmically 
divergent contribution to the pion and proton charge radii~\cite{radii}.
Lighter quark masses
will be required to check the chiral extrapolation. In section V we will
examine the effects of unquenching.

A similar comparison of $\langle z^2 \rangle $ with $\langle (x^2+y^2)/2 \rangle $
for the $\Delta^+$, using as interpolating field $J_3$
defined in Eq.~\ref{delta}, gives $\langle z^2 \rangle > \langle (x^2+y^2)/2 \rangle $
(the reverse is true for the states with
spin projection $\pm 3/2$), giving suspicion of a deformation.
However, statistical errors are larger than the signal, so that
we cannot claim to see any significant deformation.

\subsection{Three-density correlation functions}
We have analysed 30 configurations at $\kappa=0.15$ and $\kappa=0.154$. 
Since we now have to consider two relative distances the 
calculation of the three-density correlations is more demanding
even using FFT.
As explained in section II
we only compute  the diagram  with  density insertions
on different quark lines. 
However, we can check the quality of this approximation by integrating
our three-density correlation over one relative distance, and comparing
the result with the two-density correlation studied in the last subsection.
If the three-density correlation contains complete information, both
expressions should agree as per Eq.~\ref{oned}.

\begin{figure}[h]
\epsfxsize=8truecm
\epsfysize=8truecm
\mbox{\epsfbox{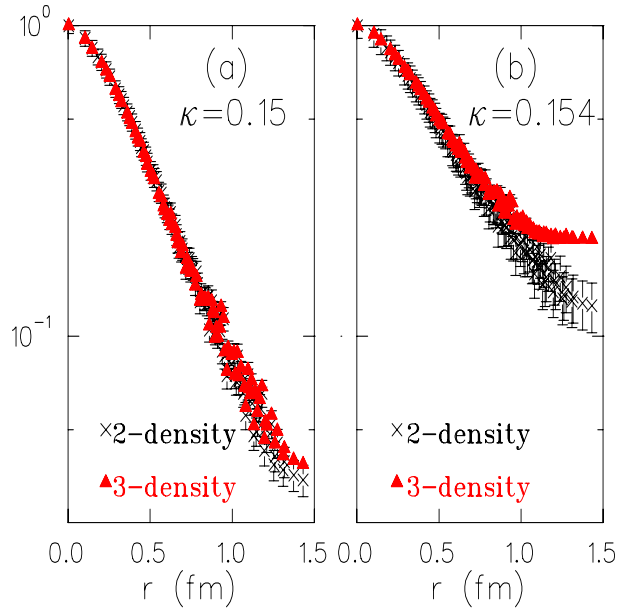}}
\vspace*{0.5cm}
\caption{Single particle density for the nucleon (a) for $\kappa=0.15$
and (b) for $\kappa=0.154$.}
\label{fig:oned}
\end{figure}

This comparison is performed in Fig.~\ref{fig:oned} for the nucleon,
for two quark masses. Although the three-density correlation, as expected,
is subject to larger finite-size effects visible for large quark separation,
both methods give virtually identical results,
indicating that we captured the dominant contribution to the three-density
correlation. A similar conclusion holds for the $\Delta$.

In any case, the usefulness of three-density correlators is to supplement 
two-density correlators and expose more detailed structure: 
single-particle observables, such as the quadrupole moment,
can be extracted directly from the two-density correlators.

\begin{figure}[h]
\epsfxsize=8.0truecm
\epsfysize=8.truecm
\mbox{\epsfbox{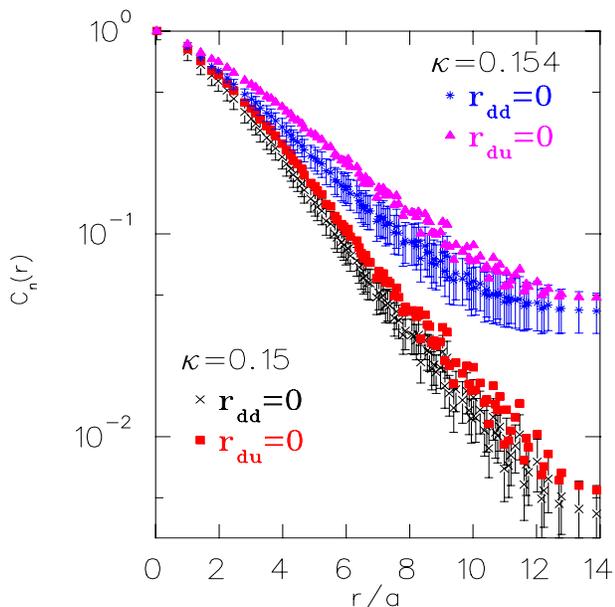}}
\caption{$u$- and $d$- quark spatial distributions in the neutron for 
$\kappa=0.15$ and $\kappa=0.154$. The errors bars for the $r_{du}=0$ case are
comparable to those for the $r_{dd}=0$ case and are omitted for clarity.}
\label{fig:neutron}
\end{figure}

The spatial $u$- and $d$- quark distributions in the nucleon can be investigated
by fixing the relative position of the other two quarks. 
We note that, since we are using degenerate $u-$ and $d-$ quarks,
 only the electric charge
differentiates the proton from the neutron. 
In Fig.~\ref{fig:neutron}
we show the spatial distributions of the $u$ and $d$ quark in the neutron, when
the relative distance between the other two quarks is fixed to zero.
For both quark masses considered,
the $d$-quark spatial distribution is slightly broader than that of the $u$-quark.
Since the total charge of the two $d$-quarks is -2/3 and that of the $u$-quark
+2/3, the broader $d$-quark spatial distribution
indicates that the charge root mean square radius of
the neutron is negative. 
For the $\Delta^+$ the two distributions are the same.
The charge radius squared  can be evaluated using
\be 
\langle r^2_{\rm ch} \rangle =\frac{ \int d^3{\bf r}_1 \int d^3{\bf r}_2 
\sum_{q=1}^3 e_q {\bf r}_q^2({\bf r}_1,{\bf r}_2) \>\> C({\bf r}_1,{\bf r}_2)}
{\int d^3{\bf r}_1 \int d^3{\bf r}_2 \>\> C({\bf r}_1,{\bf r}_2)}
\label{rms}
\ee 
where 
${\bf r}_q^2({\bf r}_1,{\bf r}_2)$ 
is the distance of each quark
 to the center of mass, in terms of the relative distances
${\bf r}_1$ and ${\bf r}_2$. Estimating  this integral
by a discrete lattice sum we obtain for the proton charge rms
$(r_p/a)^2=20 \pm 3.5 $ 
or $\sqrt{r_p^2}\sim 0.45 \pm 0.04$~fm at $\kappa=0.15$
and  $(r_p/a)^2=27 \pm 5.5 $ or $\sqrt{r_p^2}\sim 0.52 \pm 0.06$~fm at 
$\kappa=0.154$.
If we  extrapolate these two values  linearly in $m_\pi^2$ to the
chiral limit we find $\sim 0.59(4)$~fm, compared to the experimental value 
$0.81$~fm. Again we have used $a^{-1}=1.94$~GeV as determined from the
string tension  to convert to physical units. 
Using the nucleon mass in the chiral limit to set the scale
gives $a^{-1}=1.88(7)$~GeV, very close to the value obtained from
 the string tension. 
If instead we would use the rho mass to set the scale, then
$a^{-1}\approx 2.3$~GeV which gives for the proton, in the chiral limit,
a smaller value equal to   $\sqrt{r_p^2} \sim 0.50(4)$~fm.
It has  been pointed out that chiral logs can affect the
chiral extrapolation of the radii and increase their value~\cite{radii}. 
To check whether the chiral logs will produce larger values closer to the 
experimental result, additional
$\kappa$ values closer to the chiral limit will be needed.
The neutron charge radius square $r_n^2$ comes out negative for both values
of $\kappa$ as expected. We obtain $ r_n^2/r_p^2 \sim -0.25 \pm 0.08$ 
and $-0.29 \pm 0.12$ at
$\kappa=0.15$ and $\kappa=0.154$ respectively. These values are consistent
with the experimental value of $-0.146$ albeit with large errors.
Computing as a check the same quantity for the $\Delta^+$, we obtain
zero within error bars, in agreement with our earlier observation
that  the $d$- and $u$- quark spatial distributions in the $\Delta^+$
are the same.

\begin{figure}[h]
\epsfxsize=8truecm
\epsfysize=8truecm
\mbox{\epsfbox{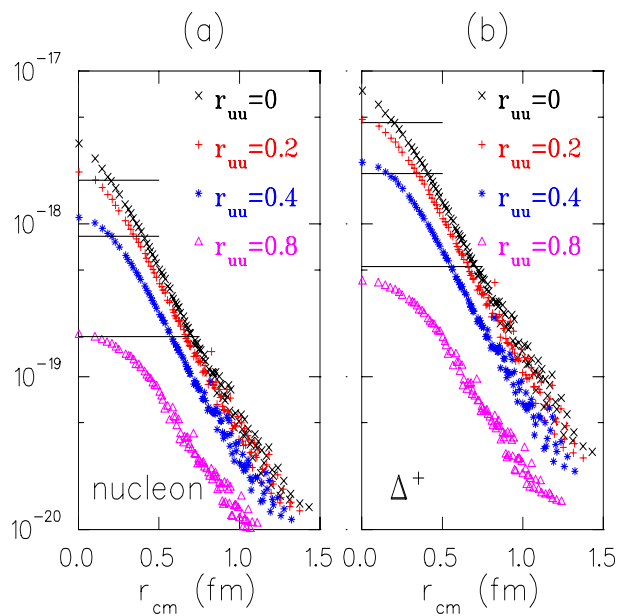}}
\vspace*{0.5cm}
\caption{$d$-quark spatial distribution with respect to the center
of mass of the two $u$-quarks, for different $u-u$ separations
(a) in the nucleon spin +1/2 projection
and (b) in the $\Delta^+$ spin +3/2 projection ($\kappa=0.15$).
The straight lines mark the value of the wave function when $r_{uu}=0$
and the $u-d$ quark separation is 0.2, 0.4 and 0.8 fm, to be compared
with $r_{cm}=0$ and $r_{uu}=0.2, 0.4$ and 0.8 fm respectively.}
\label{fig:d-distribution}
\end{figure}

Instead of fixing the relative distance between the two $u$-quarks in 
the proton or $\Delta^+$ to zero,
we can fix it to various non-zero values. In Fig.~\ref{fig:d-distribution}
we fix the $u-u$ separation, $r_{uu}$, (between 0 and $8a$ 
along a principal lattice axis),
and show the distribution of distances $r_{cm}$ between the $d$-quark and the
$u-u$ center of mass. Detailed information about nucleon structure is
contained in this figure.

In particular, let us compare configurations $(uu)-d$ where the two
$u$-quarks at at the same place i.e. $r_{uu}=0$,
and $u-d-u$ where the $d$-quark lies in the middle of the two $u$-quarks,
i.e. $r_{cm}=0$ where $r_{cm}$ is the distance from 
the $d$-quark to the center
of mass of the two $u$-quarks.
 As we will see in the next paragraph, effective models 
(both Y- and $\Delta$-Ans\"atze) predict equality of the wave functions
among these two configurations provided they have the same total size
(ie. $r_{cm}$ in $(uu)-d$ equal to $r_{uu}$ in $u-d-u$). 
Inequality reveals a finer structure than captured by effective models.
It is clear from Fig.~\ref{fig:d-distribution} that the top horizontal line
($(uu)-d$ with $r_{cm}=0.2$ fm) falls below the + sign at $r_{cm}=0$
($u-d-u$ with $r_{uu}=0.2$ fm), indicating a relative suppression of the
$(uu)-d$ configuration, i.e. a mutual $u-u$ repulsion.
The same effect is visible for a larger configuration of size 0.4 fm (second
horizontal line), but goes away or even gets inverted for yet larger 
configurations. Somewhat surprisingly, this $u-u$ repulsion at close range seems
weaker in the $\Delta^+$ where the two $u$-quarks are in the same spin state
(Fig.~\ref{fig:d-distribution}(b)) than in the nucleon where they can be
in opposite spin states (Fig.~\ref{fig:d-distribution}(a)). 
This example, where a quantitative refinement could be obtained straightforwardly
by considering lighter quarks on a larger lattice, serves to illustrate
the wealth of information contained in three-density correlators.

Having, in the non-relativistic limit, the wave function of a baryon
it is interesting to ask whether we can deduce the potential which would
yield this wavefunction. The relevant potential will of course depend
on the two relative distances between the quarks. The issue is whether
we can further reduce the degrees of freedom to effectively 
write the confining potential in terms of one distance only. Two such 
proposals exist in the literature, which make different predictions
for the linear rise of the baryonic potential.
$(i)$ The so called Y-Ansatz can be derived by a strong coupling argument~\cite{CKP}: 
the baryon potential grows in proportion to the minimal length of gluonic
string necessary to join together the three static quarks. 
The three strings join at the Steiner point.
$(ii)$ The so called $\Delta$-Ansatz is derived from a center vortex picture of
confinement~\cite{Cornwall}. The baryonic potential is simply
a sum of two-body $q\bar{q}$ potentials and grows like 
$\frac{\sigma}{2}(r_{12}+r_{23}+r_{31})$, where $\sigma$ is the $q\bar{q}$ string
tension and $r_{ij}$ the $q_i-q_j$ distance.
The difference between
these two Ans\"atze depends on the geometry of the three quark system. For
instance it vanishes when the three quarks are aligned, since then the Steiner point
which minimizes the length of flux tube joining the three quarks
coincides with the position of the middle quark. 
Unfortunately this difference never 
exceeds $\sim 15$\%, so that very accurate results 
at large quark separations $\sim 1$ fermi are needed to ascertain which
model, if any, is correct.
In a recent lattice calculation of the baryonic potential it was shown that
the $\Delta$-Ansatz is favoured for distances up to 0.8~fm
\cite{D-Ansatz}. Although this finding is in agreement
with the results of~\cite{D-Bali},
a different analysis of similar lattice results
has led others~\cite{Y-Ansatz} to the conclusion that the Y-Ansatz 
is preferred.
This issue is currently under further study both on the 
lattice~\cite{DY-Ansatz} and phenomenologically~\cite{KS}.

\begin{figure}[h]
\epsfxsize=8truecm
\epsfysize=10truecm
\mbox{\epsfbox{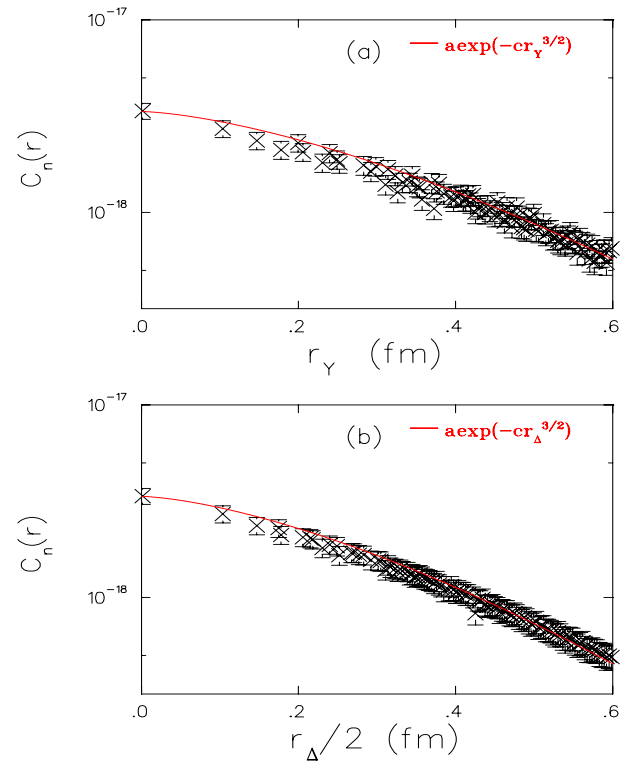}}
\vspace*{0.5cm}
\caption{Nucleon wave function (a)  versus half the $\Delta$-distance and (b)
versus the Y-distance.}
\label{fig:D-Y}
\end{figure}

\begin{figure}[h]
\epsfxsize=8.0truecm
\epsfysize=8.truecm
\mbox{\epsfbox{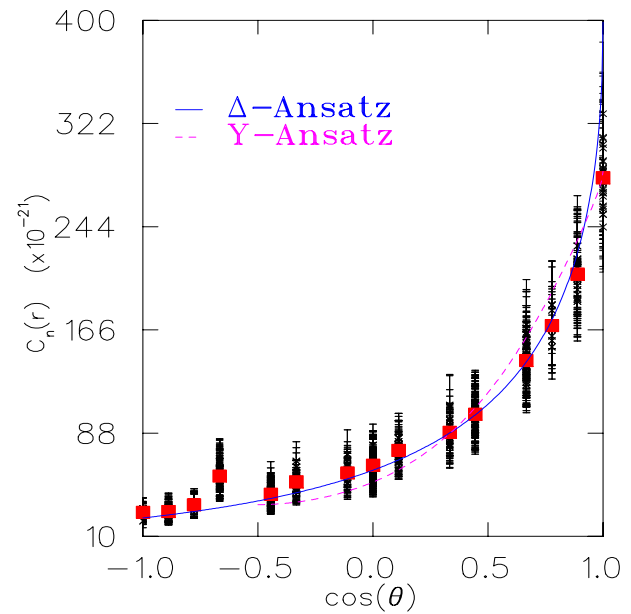}}
\vspace*{0.5cm}
\caption{Theta dependence of the nucleon wave function when the two relative
distances are fixed to $0.6 \pm 0.001$~fm. 
The filled squares are the central values
of the wave function. The solid line is the fit using the $\Delta$-Ansatz
for the angular dependence and the dashed line is the corresponding
fit using the $Y-$ Ansatz. Note that for $\cos{(\theta)}<-0.5$ the $Y$-Ansatz
coincides with the $\Delta$-Ansatz.}
\label{fig:nuc1_theta}
\end{figure}

To examine whether our wave function results favour
one of these two Ans\"atze, we plot them as a function of $r_Y$ 
(Fig.~\ref{fig:D-Y}(a)) and of $r_\Delta$ (Fig.~\ref{fig:D-Y}(b)),
where $r_Y$ is the minimal total length of the flux strings and
$r_\Delta=1/2(r_{12}+r_{23}+r_{31})$. The scatter of the data is visibly
smaller when plotted as a function of $r_\Delta$, indicating that 
$r_\Delta$ is a better effective variable than $r_Y$.
Furthermore, consider the solution to the
non-relativistic Schr\"odinger equation  with potential $V_Y\propto r_Y$
or $V_\Delta \propto r_\Delta$.
It is an Airy function which asympotically decays as $\exp(-c r^{3/2})$.    
Fits of the wave function of the nucleon to this asymptotic
form are shown in Fig.~\ref{fig:D-Y}. The fit using the $r_\Delta$ 
as the relevant distance yields
a $\chi^2/{\rm d.o.f}=0.4$ whereas using $r_Y$ 
one gets $\chi^2/{\rm d.o.f}=1.0$. Therefore both Ans\"atze provide a surprisingly 
good
description of the wave function, with a preference for the $\Delta$-Ansatz.

As a more stringent test, we fix the 
relative distances $|{\bf r}_1|=|{\bf r}_2|$ between the $d$ and the two $u$ 
quarks, and
study the wave function as a function of the $\hat{udu}$ angle, 
$\cos(\theta)={\bf r}_1 .{\bf r}_2/|{\bf r}_1| |{\bf r}_2| $,
that the relative distances
make. Both for the $\Delta$- and $Y-$ Ans\"atze the wave function should only 
depend on this angle but with a different functional dependence.
In Fig.~\ref{fig:nuc1_theta} we show the wave function for the nucleon
versus the cosine of the angle. Again the $\Delta$-Ansatz provides a better
description of the data.

\section{Density-density correlators with two dynamical quarks}
As we noted in the previous section, if the pion cloud is responsible
for hadron deformation, then the quenched and unquenched results may differ
significantly.
To investigate the
importance of dynamical quarks, we 
used the SESAM~\cite{SESAM} configurations with two dynamical degenerate 
quark species at $\beta=5.6$ on a lattice of size $16^3\times 32$. The lattice spacing
determined from the rho mass in the chiral limit
 is $a^{-1}= 2.3$~GeV~\cite{SESAM} which is  same as for
the quenched theory at $\beta=6.0$ and therefore
the physical volume is the same as in our quenched calculation.
We use this determination of the lattice spacing, which is applicable
in both the quenched and the unquenched theory,  to compare results
among both.    
We have analysed 150 configurations at $\kappa=0.156$ and 200 at
 $\kappa=0.157$.    
The ratio  of the pion mass to rho mass is 0.83 at $\kappa=0.156$ and
0.76 at $\kappa=0.157$. These values are close to the quenched mass ratios
measured at $\kappa=0.153$ (0.84) and $\kappa=0.154$ (0.78) respectively,
allowing us to make pairwise quenched-unquenched comparisons.

In Fig.~\ref{fig:wfs}(b) we plot the two-density correlators
for the four hadrons for light sea quarks ($\kappa=0.157$). Comparing with
the quenched results ($\kappa=0.154$), one sees that
the pion size remains the same, but that the rho, nucleon and $\Delta^+$
sizes increase. The $\Delta^+$ is now clearly
larger than the rho and  the rho correlator
decays more slowly than the nucleon. This slower decay is
also  visible in the quenched case for $\kappa=0.154$ and $0.155$ 
 but not for the smaller values of $\kappa$. 
The invariance of the pion size can be also seen in 
Fig.~\ref{fig:meson-radii} which shows the second moments.
 This result is consistent with
calculations in the Dyson-Schwinger framework
which concluded that the pion cloud, consisting of physical pions,
contributes only 15\% to the root mean square radius of the 
pion~\cite{Alkofer}. In our simulations the effect is
expected to be even smaller since the pion mass is larger than 600~MeV.
On the other hand, Fig.~\ref{fig:meson-radii} shows that unquenching increases the rho size and therefore the ratio
of the rho charge radius to the pion's approaches the 
experimental value. This ratio is expected to be
 more reliable than the individual moments since lattice artifacts
partly cancel out.

\begin{figure}[h]
\epsfxsize=8.0truecm
\epsfysize=11.5truecm
\mbox{\epsfbox{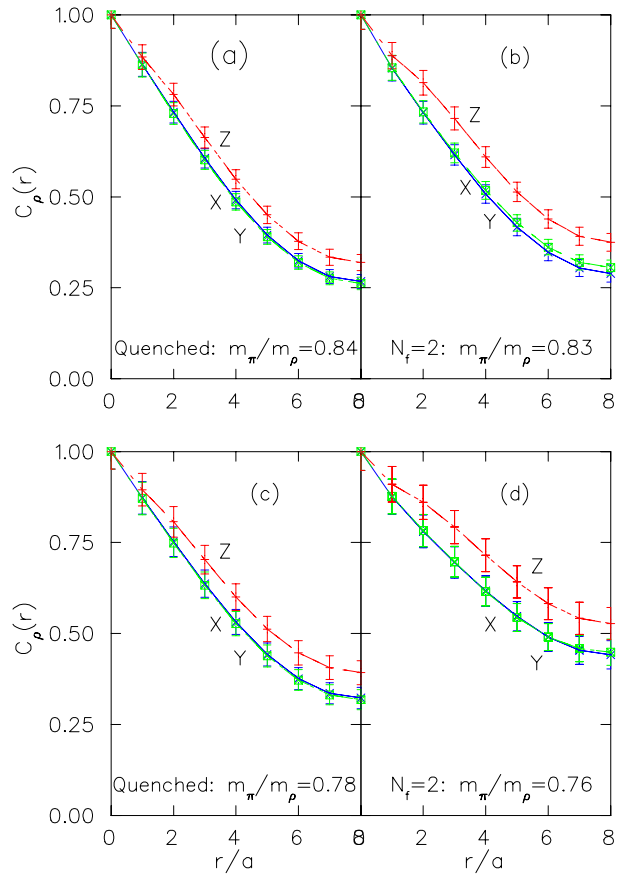}}
\vspace*{0.5cm}
\caption{Asymmetry for the rho wave function 
 (a) in the quenched approximation  at $\kappa=0.153$ and (c) at $\kappa=0.154$
and (b) for two dynamical quarks at $\kappa=0.156$ and (d) $\kappa=0.157$ .}
\label{fig:SESAM rho}
\end{figure}

 From Fig.~\ref{fig:SESAM rho} we compare
the quenched and unquenched asymmetry in  the rho for comparable ratios
$m_\pi/m_\rho$. 
The main observation is that the asymmetry in
the rho grows in full QCD. This is clearly seen in the (a)-(b) comparison,
i.e. for the heavier dynamical quarks.
For the lighter quarks (c)-(d), this growth in the asymmetry is still there  
but the effect seems less pronounced. On the other hand, finite-size
effects are more important. Disentangling the two would require larger lattices.

\begin{figure}[h]
\begin{minipage}{7.5cm}
\epsfxsize=7.5truecm
\epsfysize=7.5truecm
\mbox{\epsfbox{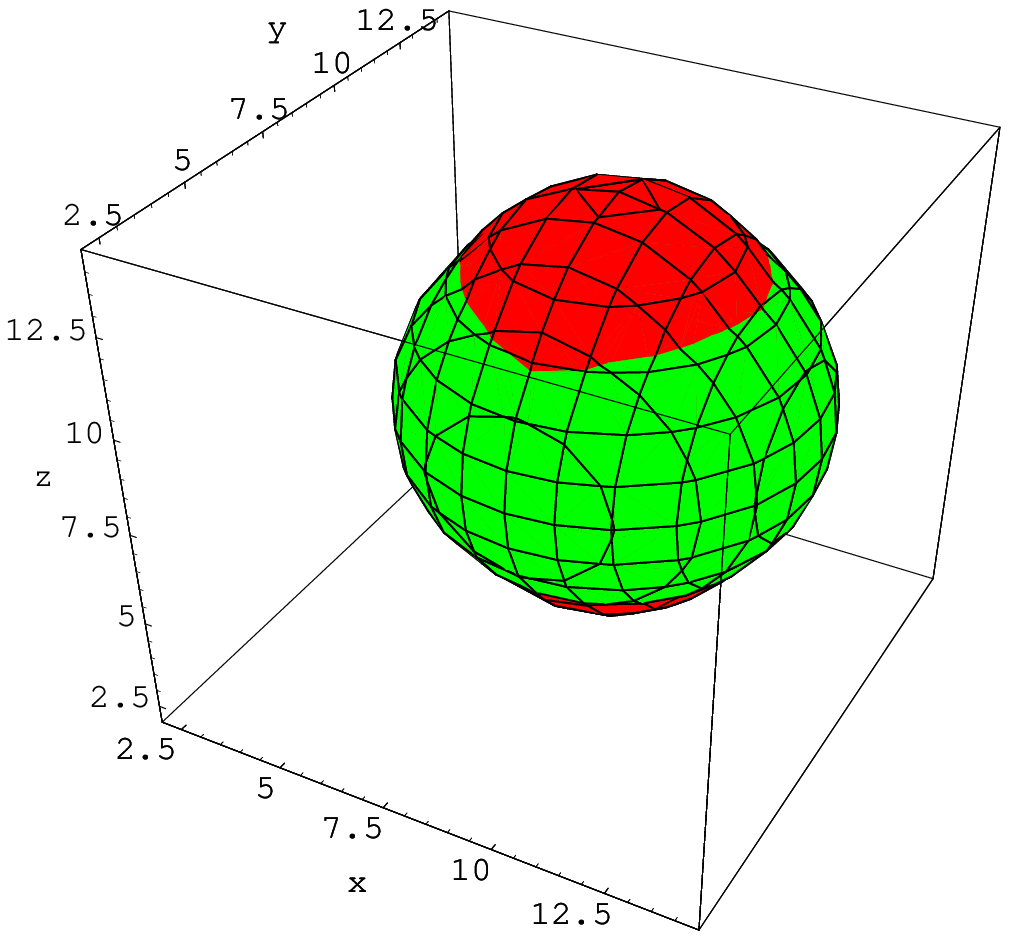}}
\end{minipage}
\begin{minipage}{7.5cm}
\epsfxsize=7.5truecm
\epsfysize=7.5truecm
\mbox{\epsfbox{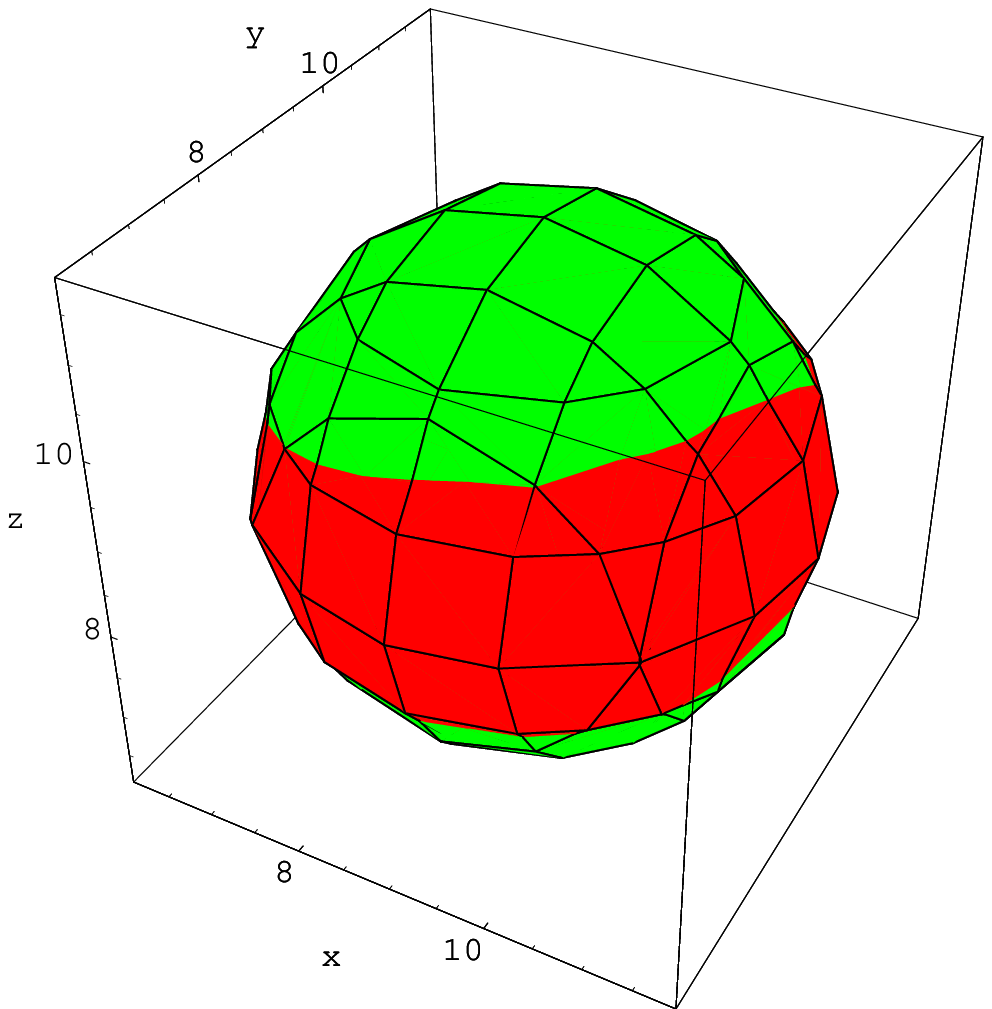}}
\end{minipage}
\vspace*{0.5cm}
\caption{Three-dimensional contour plot of the correlator (black):
upper for the rho state with 0 spin projection (cigar shape)
 and  lower for the $\Delta^+$ state with 
+3/2 (slightly oblate) spin projection
for two dynamical quarks at $\kappa=0.156$.
Values of the correlator (0.5 for the rho, 0.8 for the $\Delta^+$) were chosen 
to show large distances but avoid finite-size effects.
We have included for comparison
 the contour of a sphere (grey).}
\label{fig:3D-contour}
\end{figure}

The same analysis for the $\Delta^+$ gives no
definite results regarding the deformation. Whereas in all channels we 
do observe a deformation obeying the
sign relations obtained in Section II with the Wigner-Eckart theorem,
the deformation remains  small with large statistical errors.
 However it is interesting to look at a three-dimensional contour
plot for the $\Delta^+$ in the $+3/2$ spin state and compare it
to the rho 0-spin state. In Fig.~\ref{fig:3D-contour} we show these
contour plots for  the heavier sea quarks that we analyzed. The elongation
in the rho is clearly visible whereas the $\Delta^+$ appears to be
squeezed. Note that squeezing in this channel
means that the unphysical channel using an interpolating field
 a $J_3$ in Eq.~\ref{delta}
is elongated. The statistical uncertainties 
are larger
for the  lighter sea
quark but the trend for the deformation remains the same.
The overall conclusion is that unquenching
 seems to increase the deformation for the $\rho$ and the $\Delta$,
which would imply that the pion cloud contributes to the deformation
of these hadrons. However more statistics, a bigger lattice and lighter
quark masses will be needed to consolidate this observation.

\section{Conclusions}
Two-density correlators for mesons and baryons are shown to contain
rich information on their structure. 
For the baryons these correlators reduce to the one-particle density 
in the non-relativistic limit.

The overall quark mass dependence
of the correlators for the pion, the rho, the nucleon and
the $\Delta^+$ is found to be rather weak. The rho correlator
shows the strongest dependence on the quark mass, and the nucleon and the $\Delta$ 
the weakest. For the quark masses that we have studied in this work
unquenching has the strongest effect on hadron sizes
for the rho and the $\Delta$,
and the weakest for the pion and the  nucleon.
In the quenched approximation we find that the rho state with 0-spin projection
 is prolate,
whereas for the $\Delta^+$ no statistically significant 
deformation is seen. Since for the rho there
are no virtual pions from backward moving quarks 
 in the quenched approximation,
the rho deformation is a dynamic property of quarks forming a vector particle.
Adding sea quarks while keeping the pion to rho mass ratio constant,
we observed an increase in the rho deformation,
and a slight oblate deformation of the $\Delta^+$ $+3/2$ spin state.
The pion cloud may be
responsible for this effect. As already pointed out,
lighter sea quark masses together with a larger lattice to well contain
the hadrons will be needed in order to study further the role
of the pion cloud. 

Three-density correlators for
baryons, although computed neglecting insertions on the same quark line,
are shown to reproduce the two-density correlators 
when integrated over one relative coordinate.
Therefore one expects the contribution
of the neglected diagram with two insertions on the same quark line to be small. 
 With the three-density
correlators  baryonic structure can be explored in more detail.
In the neutron we
 clearly detect a broader $d$-quark spatial distribution compared
to that of the $u$-quark. This accounts for the negative charge 
square radius of the neutron 
observed experimentally. 
By comparing the $u$- and $d$- spatial distributions
in the proton and in the $\Delta^+$ we observe that there is a 
 preference for the two $u$-quarks  
to be at 180$^0$
rather than at the same place. More statistics are needed
to consolidate the trend observed here.

Information on the baryonic potential can be extracted from fits
to the three-density correlators of the nucleon. 
By performing a fit to the radial and angular
dependence we find that the baryonic wave functions are
better described by a confining potential which is the sum of two-body
potentials known as the $\Delta$-Ansatz, at least for relative distances
of $\sim1$~fm that we can probe in this work.

\vspace*{0.5cm}

\noindent
\underline{Acknowledgements:} We thank C. N. Papanicolas for encouraging
us to look into the issue of deformation and for discussions.
The $SU(3)$ $16^3\times 32$ quenched 
lattice configurations were obtained from the
Gauge Connection archive~\cite{NERSC}. We thank the SESAM collaboration
for giving us access to their  dynamical lattice configurations.

\end{document}